\documentclass[11pt,a4paper]{article}
\usepackage{graphics}
\usepackage{graphicx}
\usepackage{amssymb}
\usepackage{cite}
\usepackage{verbatim} 
\usepackage[usenames]{color}

\newcommand{\bEq}{\begin{equation}}
\newcommand{\eEq}{\end{equation}}
\newcommand{\bEQ}[1]{\begin{equation} \begin{array}{#1}}
\newcommand{\eEQ}{\end{array} \end{equation}}

\newcommand{\correction}[1]{\textcolor{black}{#1}}

\begin{document}

\pagestyle{empty}

\null

\vfill

\begin{center}

{\Large  
{\bf On the magnitude of error in determination of rotation axes}}\\

\vskip 1.0cm
A. Morawiec
\vskip 0.5cm 
{Institute of Metallurgy and Materials Science, 
Polish Academy of Sciences, \\ ul. Reymonta 25, 30-59 Krak{\'o}w, Poland.
}
\vskip 0.5cm 
E-mail: nmmorawi@cyf-kr.edu.pl \\
Tel.: ++48--122952854, \ \ \  Fax: ++48--122952804 \\

 \end{center}

\vfill

\noindent
{\bf Synopsis}\\
Assuming that errors affecting crystal rotations are random, 
formulas for the average and maximum errors of rotation axes are provided.

\vfill

\noindent
{\bf Abstract}\\
Rotation axes (together with rotation angles)
are often used to describe crystal orientations and misorientations,
and they are also needed to characterize some properties of crystalline materials.
Since experimental orientation data are subject to errors,
directions of axes obtained from such data are also inaccurate. 
A natural question arises: given a resolution of input rotations, 
what is the average error of the rotation axes?
\correction{Assuming that rotation data characterized by rotation angle $\omega$ 
deviate from actual data by error rotations with fixed angle $\delta$ but otherwise random, 
the average error of rotation axes of the data is expressed analytically 
as a function of $\omega$ and $\delta$.}
%
%
Moreover, a scheme for using this formula in practical cases 
when rotation errors \correction{$\delta$} follow the von Mises-Fisher distribution is described.
Finally, the impact of crystal symmetry on 
determination of average errors of axis directions is discussed. 
The presented results are important for assessing the reliability of rotation axes 
in studies where directions of crystal rotations play a role, 
e.g. in identifying deformation slip mechanisms.

\vskip 0.5cm

\noindent
\textbf{Keywords:} crystal orientation; misorientation; rotation axis; crystallographic texture; 
slip system \\

\vskip 0.2cm

\noindent
\hfill \today

\newpage

\pagestyle{plain}

\noindent
\section{Introduction}
Various rotation parameterizations are used to describe orientations and 
misorientations of crystllites. 
Texture calculations often rely on parameters that are difficult to interpret directly, but 
in interpersonal communication, orientations and misorientations 
are usually expressed in terms of easily interpreted 
rotation axes and rotation angles.
A rotation axis consists of points invariant under the rotation.
Knowing axis direction may give insight into physics of the process causing the rotation. 
Directions of rotation axes are explicitly specified to determine some material characteristics.
For instance, they are key to establishing 
slip systems active during plastic deformation 
or, more generally, to investigating cumulative motions of dislocations 
leading to macroscopic rotations of crystal lattices.
Also, criteria for classifying grain boundaries are based on directions of rotation axes.
Some orientation distributions are axial, 
i.e., crystal orientations are related by rotations about a specific axis.
The issue of determining and interpreting 
rotation axes often appears in studies on crystallographic textures; see, e.g.,
\cite{Prior_1999,Cross_2003,Reddy_2005,Chun_2010,Yamasaki_2013,Jeyaraam_2019,Li_2022}.

Experimental orientation data are subject to errors.
The magnitudes of errors 
are characteristic of
individual orientation measurement techniques.
The orientation resolution 
typically ranges 
from hundredths of a degree to several degrees.
For standard EBSD orientation mappings, 
it is usually claimed to be about $0.5^{\circ}$--$1.0^{\circ}$
\cite{Godfrey_2002,Bate_2005,Morawiec_2014}.
\correction{Tilting a crystal by several degrees may be needed to 
detect a change in a TEM spot diffraction pattern. 
Magnitudes of errors for some non-diffraction techniques can be quite large;
the orientation resolution of the 
metallographic etch-pitting technique is of the order of $10^{\circ}$.	
}

\correction{
For simplicity, the text below refers to errors 
affecting single crystal orientation measurements, but it can also be 
reformulated in terms of features of textures.
Orientation data obtained from polycrystalline materials are usually 
reduced to texture components with component-defining central orientations and some component spreads.
The spread of a component can be viewed as a measure of 
deviations from the central orientation.}

The rotation axes obtained from error-affected orientation data are also inaccurate,
and inferences based on such axes become uncertain. 
The question arises about magnitudes of errors in determining axis directions. 
This problem is the subject of consideration below.
A correction is made to the  hitherto used expression, 
originally presented as the formula for the average error in axis direction.
In parallel, a formula for the maximum error in axis direction is given.
Moreover, a scheme is presented for estimating the average error in axis direction 
in the case where the rotation errors have the spherical von Mises-Fisher distribution.
Since symmetry complicates the determination of axis directions,
the opportunity is taken to  clarify the issue of 
the average error in axis direction for
rotations representing orientations and misorientations of symmetric crystals.

The remainder of this paper is organized as follows. 
The next section recalls some basic facts about 
rotations.
Section \ref{sect_no_symm} discusses average
deviations of rotation axes for rotation errors of fixed magnitude.
Then, in section \ref{sec:distributed_delta}, the case of distributed magnitudes 
of rotation errors  is briefly considered. 
The issue of the influence of crystal symmetry on determining 
the average error of rotation axis is discussed in section \ref{sec:symmetric_case}.

\section{Basics \label{sect_Basics}}

Throughout this paper, it is assumed that the crystal point group contains inversion,
and considerations are limited to proper rotations.
Orientation of a crystal is described by a rotation that transforms 
a crystal reference frame into an established sample reference frame.
The misorientation between two crystals 
is represented by a rotation that transforms 
the reference frame of the first crystal into 
the frame of the second crystal.

\correction{Interest in axes of rotations is often associated with research on 
	misorientations between crystals of the same type.
	The axis and angle of the rotation associated with a misorientation 
	are usually referred to as 
	misorientation axis and misorientation angle, respectively.
	Misorientation axes are usually specified with respect to 
	the crystal reference frame by indices of their directions.}

\correction{In the case of orientations, axis directions are important 
	when orientation-dependent changes 
	need to be coupled to external forces, stimuli or constraints.
	(E.g., for understanding plastic deformation of polycrystals, 
	it is important to know the link between axes of 
	grain rotations and stresses applied to the specimen.)
	The rotation axes for orientations are frequently specified with respect to 
	the sample reference frame.}

\correction{As a side note, it is worth mentioning 
	misorientations between crystals of different phases.
	Interpretation of such misorientations is complicated by the fact that
	a change of the setting of the crystal coordinate systems in one of the phases, 
	changes the null misorientation, i.e., the latter is not unique.
	However, as in the case of single phase materials, misorientation axes have a 
	the same fundamental meaning:
	when one crystal  rotates with respect to the other,
	the axis of rotation is along the direction invariant in both crystals.}

With $\mathbf{k}$  being a unit vector along the rotation axis
and $\omega$ denoting the rotation angle, 
the pair $(\mathbf{k},\omega)$ uniquely identifies the right-hand rotation.  
It is clear that with $n$ denoting an integer, $(\mathbf{k},\omega)$, $(\mathbf{k},\omega + 2 n \pi)$ 
and $(-\mathbf{k},-\omega + 2 n \pi)$ represent the same rotation. 
To limit these ambiguities, the domain of the rotation parameters needs to be specified.
The most convenient, natural and commonly used is the one which 
consists of rotations closest to the null rotation $\mathit{I}$
(i.e., the rotation by the angle $\omega=0$).
With this domain, the angles of rotations are non-negative
and do not exceed $\pi$, 
and the vectors $\mathbf{k}$ cover the complete unit sphere.
In the case of half-turns, the axes are still ambiguous: $(\mathbf{k},\pi)$  and $(-\mathbf{k},\pi)$
represent the same rotation.
The parameterization by $(\mathbf{k},\omega)$ is also singular at the point $\mathit{I}$, for which
the axis is arbitrary.\footnote{Some rotation parameterizations are regular at $\mathit{I}$
(e.g., the parameterization by the rotation vector $\omega \mathbf{k}$).
It is known, however, that 
there is no global singularity--free parameterization of rotations \cite{Stuelpnagel_1964},  
i.e., the pair $(\mathbf{k},\omega)$ is in this respect 
no different from other parameterizations.} 
The point $\mathit{I}$ corresponds to the case of no rotation and is ignored here;
it is assumed that angles of 
all rotations are positive, i.e., $0 < \omega \leq \pi$. 
The choice of the domain of the parameters $(\mathbf{k},\omega)$ is essential for 
the considerations presented below.
It is assumed that all analyzed axis/angle parameters are in the described domain.

The deviation $\beta$ between axes along $\mathbf{k}$ and $\mathbf{k}'$
is defined as the angle between these vectors, i.e., 
$\beta =\arccos(\mathbf{k} \cdot \mathbf{k}')$. 
Angles appearing in theoretical expressions are in radians,
while in practical circumstances, angles are expressed in degrees.

\section{Axis deviations for rotation errors of given magnitude \label{sect_no_symm}}

Let $\delta$ denote the angular distance between 
the  measured (i.e., approximate) rotation \correction{$g' \sim (\mathbf{k}',\omega')$} and 
the true rotation    \correction{$g \sim (\mathbf{k},\omega)$}
of an object with no symmetry other than 
invariance with respect to inversion.
With the special orthogonal matrices representing the rotations denoted by the same symbols 
as the rotations themselves, 
one has 
$\delta = \arccos( (\mbox{Tr} (g' g^{-1}) - 1)/2 )$.

\begin{figure}
	\begin{picture}(300,220)(0,0)
		\put(10,100){\resizebox{4.0 cm}{!}{\includegraphics{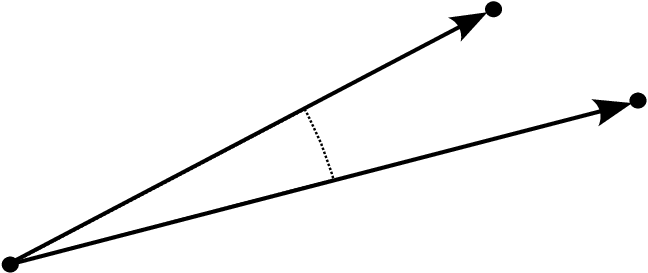}}}
		\put(180,0){\resizebox{8.5 cm}{!}{\includegraphics{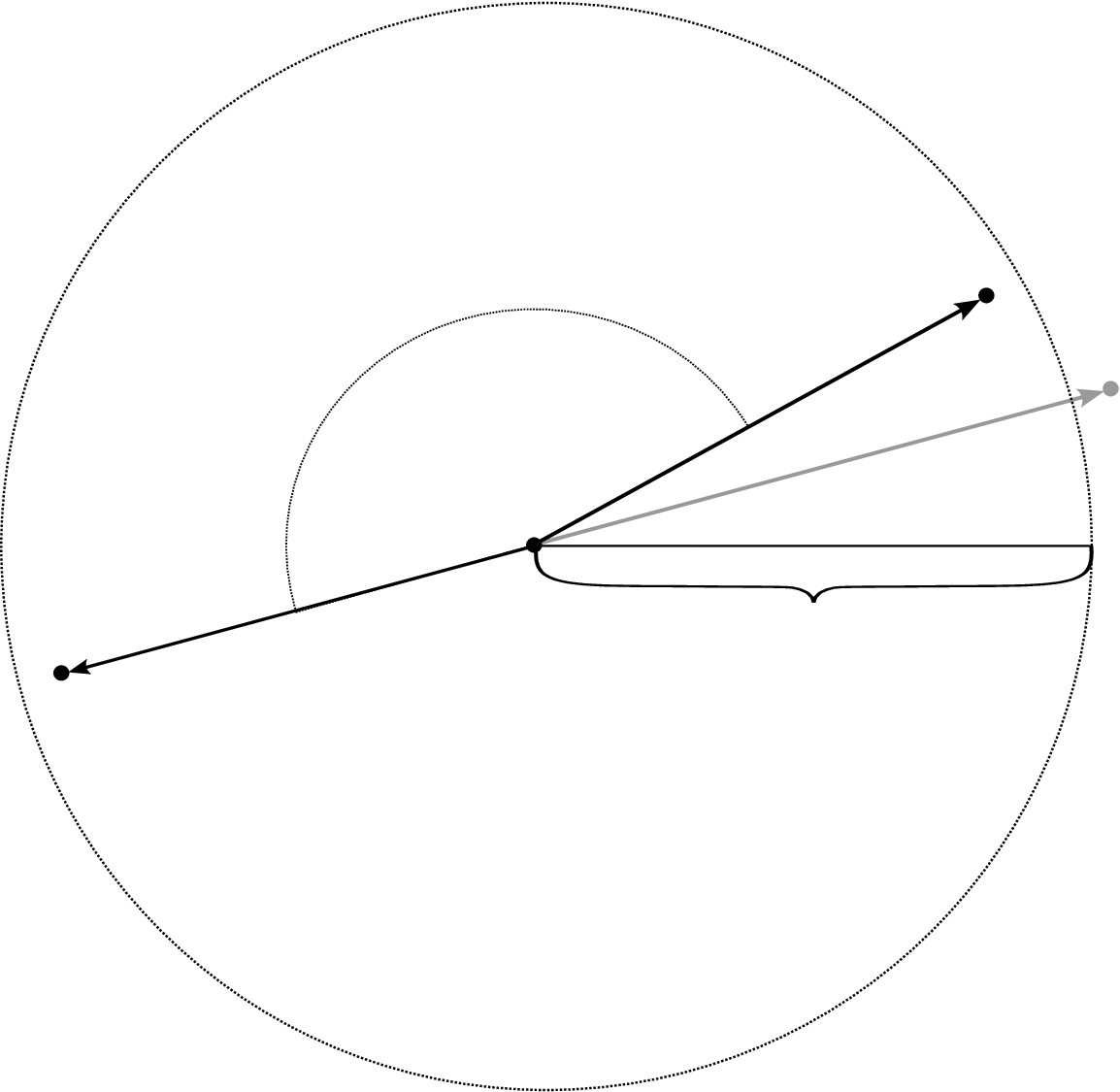}}}
		
		\put(0,170){\textit{a}}
		\put(7,104){$\mathit{I}$}
		\put(66,139){$\omega \mathbf{k}$ }
		\put(85,112){$\omega' \mathbf{k}'$ }
		\put(70,122){$\beta$}
		\put(100,145){$g$}
		\put(126,128){$g'$}
		
\put(180,210){\textit{b}}
\put(290,120){$\mathit{I}$ }
\put(275,153){$\beta$}
\put(355,160){$\omega \mathbf{k}$ }
\put(213,86){$\omega' \mathbf{k}'$ }
\put(360,127){$(\omega' -2 \pi) \mathbf{k}'$ }
\put(423,150){$g'$}
\put(187,94){$g'$}
\put(395,165){$g$}
\put(353,97){$\pi$}
	\end{picture}
	\vskip 0.0cm
	\caption{(\textit{a}) Scheme illustrating definition of the angular deviation $\beta$ between axes of rotations.
		The true rotation $g$ and 
		the deviated rotation $g'$ are represented by the rotation vectors $\omega \mathbf{k}$ and $\omega' \mathbf{k}'$,
		respectively. 
	(\textit{b}) Illustration of the case with $\omega \geq \pi-\delta$.
	The rotation $g'$ corresponds to 
	$\omega' \mathbf{k}'=(-\omega') (-\mathbf{k}')$ and to
	$(-\omega' + 2 \pi) (-\mathbf{k}')=(\omega' - 2 \pi) \mathbf{k}'$, but the latter point  
	is outside the domain adopted for the rotation vectors.
	}
	\label{Fig_nearPi}
\end{figure}
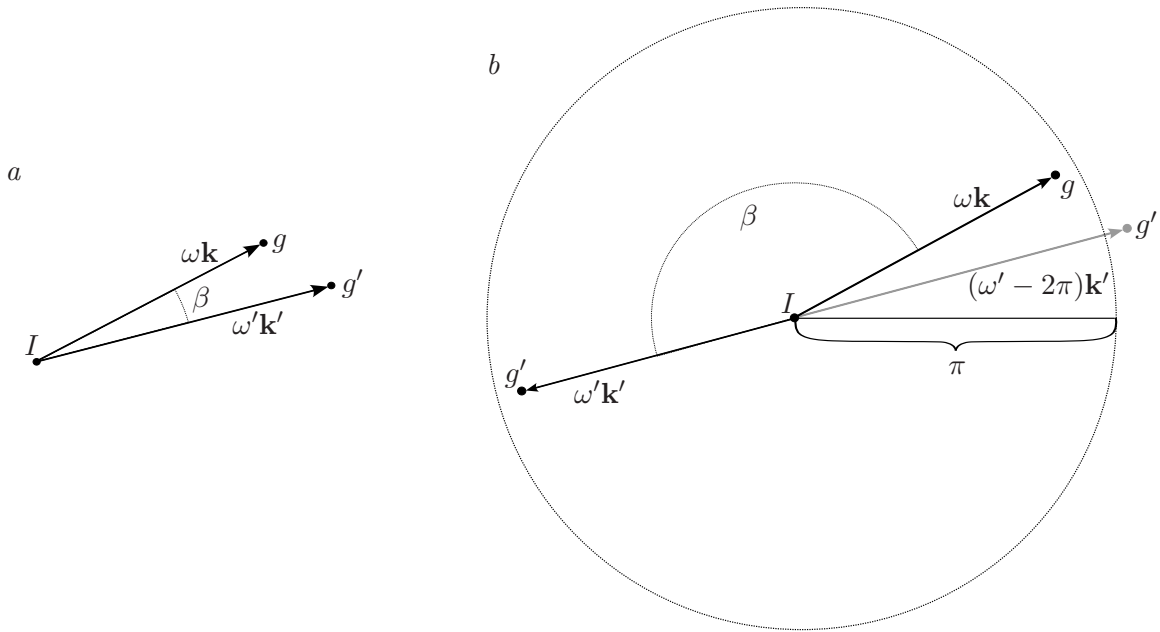

Given a fixed 
$\delta$ ($0< \delta \leq \pi$), 
the following formal question arises: assuming the error rotations
\correction{$g' g^{-1} \sim (\mathbf{h},\delta)$}
have random axes,
what is the average of the angle $\beta$ between the axes of $g$ and $g'$. 
\correction{
With vectors $\mathbf{h}$ uniformly distributed over the unit sphere $S^2$ (and
the remaining parameters fixed), the angle $\beta$ 
depends on $\mathbf{h}$, i.e., $\beta = \beta(\mathbf{h})$, and the 
average  of $\beta$ is $\langle \beta \rangle =
\int_{S^2} \beta(\mathbf{h}) \, \mbox{d}\mathbf{h}$. }
Clearly, $\langle \beta \rangle$ must depend on the rotation angle $\omega$ of $g$. 
For $\omega$ near zero, the value of 
$\langle \beta \rangle$ is expected to be large.
Moreover,  when $g$ is nearly a half-turn, an error may cause 
a large change of axis direction (Fig.~\ref{Fig_nearPi}).

An analytically derived expression intended to answer the above question was given 
by Bate et al. \cite{Bate_2005}.
They claim that, on average, the axis of the error-affected rotation is 
inclined to the true axis by the angle $\arctan(\delta/\omega)$.
This formula is often referred to in research papers 
(e.g., 
\cite{Wilkinson_2006, Brough_2006, Humphreys_2007, Farooq_2008,	Quey_2010, Wilkinson_2010, Gardner_2011, Ram_2015}) 
and PhD theses (e.g., \cite{Albou_2010b,Renversade_2016,Qu_2023}). 
However, when deriving the above expression, 
it was incorrectly assumed 
that the average of $\beta$ as a function of a variable is 
equal to the value of the function at the average of the variable. 
Therefore,  the formula of Bate et al. 
does not represent the angle $\langle \beta \rangle$.
Moreover, the paper 
\cite{Bate_2005} ignores 
the case of $\omega \geq \pi-\delta$.

Assuming the error rotations have random axes,
it can be shown \correction{(see Appendix A)} that 
with $\delta < \omega < \pi-\delta$, 
the average of $\beta$ is given by
\bEq
\langle \beta \rangle(\omega,\delta) =
\frac{\pi}{2} \ 
\frac{  
	\tan \left( \delta/4 \right)}{\sin \left(\omega/2\right)} \ .
\label{eq_beta_average} 
\eEq
There is also a closed-form expression for $\langle \beta \rangle$ 
\correction{covering the entire range}
$0 <  \omega \leq \pi$, 
but it is much more complex than (\ref{eq_beta_average}), so it is 
listed in Appendix A. 

Closely related to $\langle \beta \rangle$ is 
the maximal angle $\beta_{\max}$ between axes of the rotations 
$g$ and $g'$.
If $\delta < \omega < \pi-\delta$, 
the maximal deviation is  
\bEq
\beta_{\max}(\omega,\delta) = \arccos \left(  \sqrt{\frac{\cos \delta-\cos \omega}{1 -\cos\omega}}\right)  
\label{eq_beta_max} \ .
\eEq
If $\omega  \leq \delta$ or 
$\omega \geq \pi-\delta$, then $\beta_{\max} = \pi$. 
The justification for (\ref{eq_beta_max}) is provided in Appendix A.

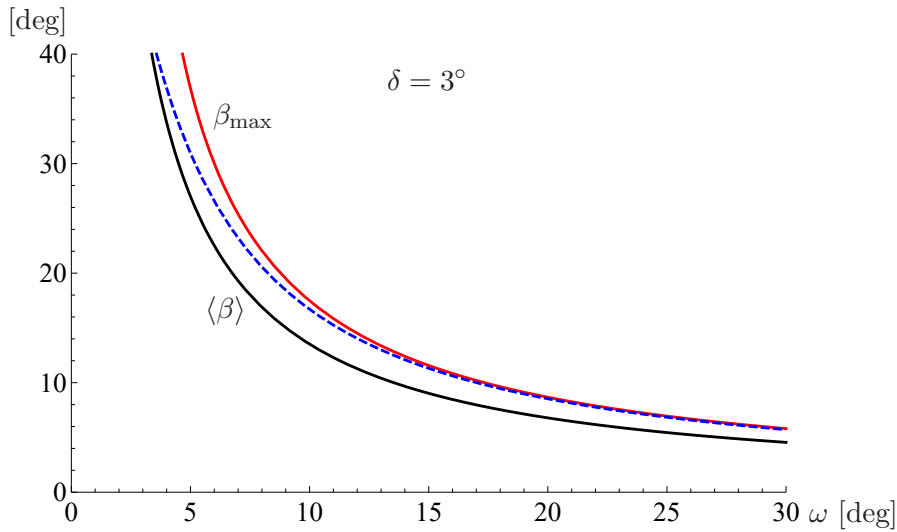
\begin{figure}[h]
	\begin{picture}(300,195)(0,0)
		\put(60,0){\resizebox{10.0 cm}{!}{\includegraphics{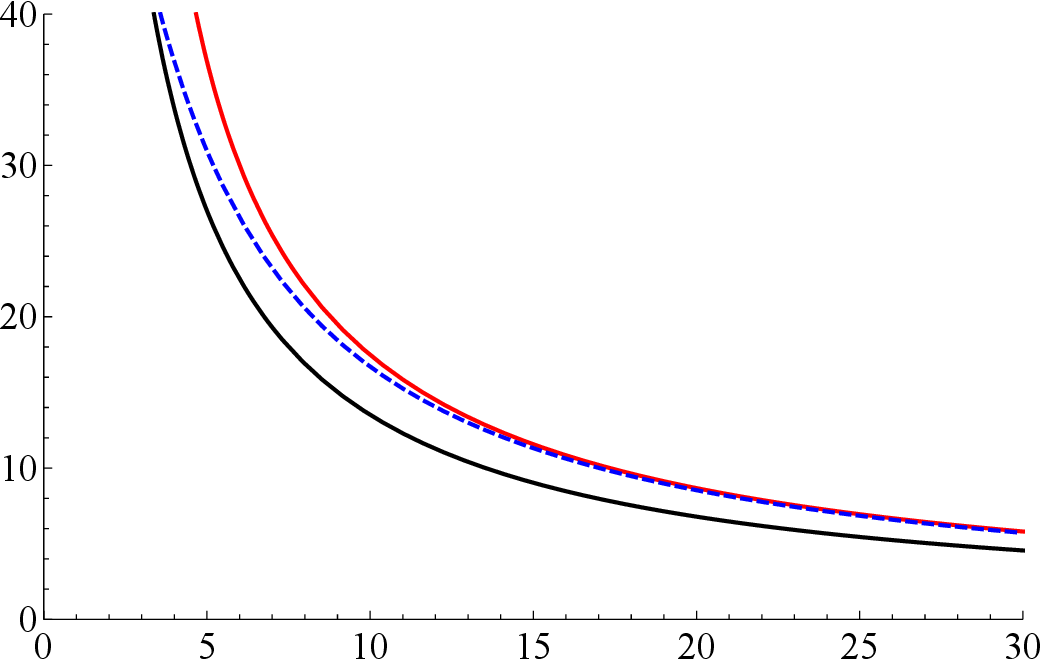}}}
		
		\put(348,0){$\omega$ [deg]}
		\put(48,186){[deg]}
		\put(122,77){$\langle \beta \rangle$}
		\put(125,150){$\beta_{\max}$}
		\put(190,163){$\delta=3^{\circ}$}
	\end{picture}
	\vskip 0.0cm
	\caption{The average $\langle \beta \rangle$ and maximal 
		$\beta_{\max}$ angles between axes of the actual and an error-affected rotations
		versus the rotation angle $\omega$ for errors of magnitude $\delta=3^{\circ}$. 
		The dashed blue curve corresponds to the formula of Bate et al. \cite{Bate_2005}.
	}
	\label{Fig_Bate_0}
\end{figure}

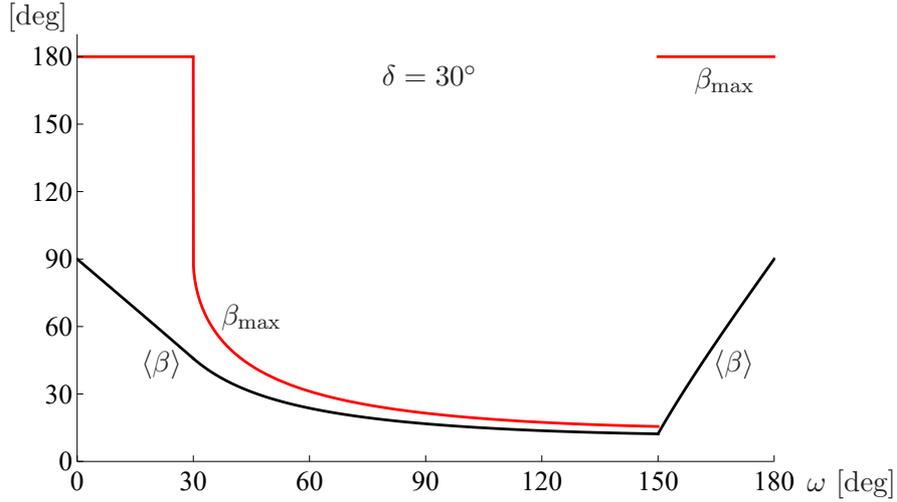
\begin{figure}[h]
	\begin{picture}(300,185)(0,0)
		\put(60,0){\resizebox{10.0 cm}{!}{\includegraphics{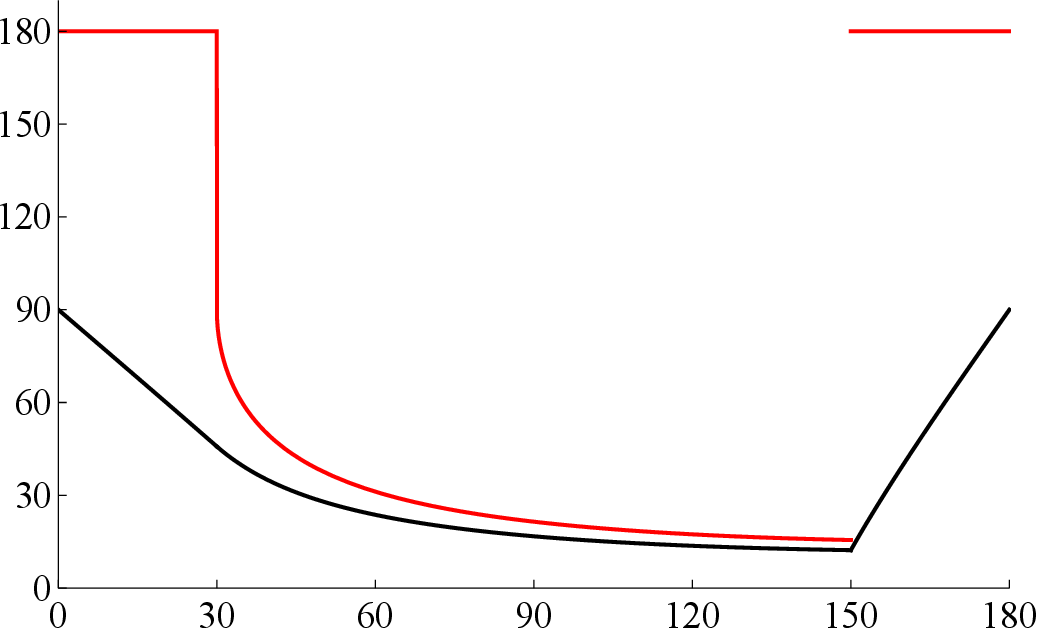}}}
		
		\put(349,0){$\omega$ [deg]}
		\put(100,45){$\langle \beta \rangle$}
		\put(314,45){$\langle \beta \rangle$}
		\put(130,63){$\beta_{\max}$}
		\put(307,152){$\beta_{\max}$}
		\put(190,153){$\delta=30^{\circ}$}
		\put(50,176){[deg]}
    \end{picture}
	\vskip 0.0cm
	\caption{The average $\langle \beta \rangle$ and maximal $\beta_{\max}$
		angles between axes of the actual and error-affected rotations
		versus rotation angle $\omega$ for errors of magnitude $\delta=30^{\circ}$. 
		The plot of $\langle \beta \rangle$ for $0< \omega \leq \pi$ is drawn using  
		formula listed in Appendix A. 
	}
	\label{Fig_Delta_30deg}
\end{figure}

Example plots of the dependence of  
$\langle \beta \rangle$ and $\beta_{\max}$ 
on $\omega$ for fixed $\delta$ $(=3^{\circ})$ are shown in
Fig. \ref{Fig_Bate_0}. 
The nature of $\langle \beta \rangle$  and $\beta_{\max}$ as functions 
of $\omega$ in the domain between $0$ and $\pi$
is better seen in plots for larger $\delta$; 
an example of such a plot is in Fig. \ref{Fig_Delta_30deg}.
Values of $\langle \beta \rangle$ and $\beta_{\max}$
for selected small error magnitudes are listed in Table~\ref{tab_betas}.
For other numerical estimates of $\langle \beta \rangle$ and $\beta_{\max}$, 
see \cite{Wilkinson_2001,Tong_2022}.

Clearly, inferences based on axis directions characterized by large 
$\langle \beta \rangle$ and $\beta_{\max}$  must be avoided.
Nothing can be done if $\omega$ is close to $0$;
in this case, axis directions are unreliable.
Data with $\omega$ close to $\pi$ can be used by taking into account 
both the obtained axis direction and the opposite direction.

\begin{table}
	\begin{center}
		\begin{tabular}{r|rr|rr|rr}
         & \multicolumn{2}{c|}{ $\delta=0.5$} 
         & \multicolumn{2}{c|}{ $\delta=1.0$} 
         & \multicolumn{2}{c}{ $\delta=3.0$} \\        
\hline		
$\omega$ & $\langle \beta \rangle$ & $\beta_{\max}$ & $\langle \beta \rangle$ & $\beta_{\max}$ 
& $\langle \beta \rangle$ & $\beta_{\max}$ \\
\hline		
	1.0   & 22.50 & 30.00 & 45.00 & 180.00 & 75.00 & 180.00 \\
	2.0   & 11.25 & 14.48 & 22.50 &  30.00 & 60.01 & 180.00 \\
	3.0   &  7.50 &  9.60 & 15.00 &  19.47 & 45.01 & 180.00 \\
	5.0   &  4.50 &  5.74 &  9.00 &  11.54 & 27.01 &  36.88 \\
	7.0   &  3.22 &  4.10 &  6.43 &   8.22 & 19.30 &  25.39 \\
	10.0  &  2.25 &  2.87 &  4.51 &   5.75 & 13.52 &  17.48 \\
	15.0  &  1.50 &  1.92 &  3.01 &   3.83 &  9.03 &  11.57 \\
	30.0  &  0.76 &  0.97 &  1.52 &   1.93 &  4.55 &   5.80 \\
	60.0  &  0.39 &  0.50 &  0.79 &   1.00 &  2.36 &   3.00 \\
	90.0  &  0.28 &  0.35 &  0.56 &   0.71 &  1.67 &   2.12 \\
	150.0 &  0.20 &  0.26 &  0.41 &   0.52 &  1.22 &   1.55 \\
	177.0 &  0.20 &  0.25 &  0.39 &   0.50 &  1.18 & 180.00 \\
	178.0 &  0.20 &  0.25 &  0.39 &   0.50 & 30.93 & 180.00 \\
	179.0 &  0.20 &  0.25 &  0.39 & 180.00 & 60.50 & 180.00 \\
\end{tabular}
\end{center}
\caption{Numerical values of average and maximal 
	angles between axes of the actual and error-affected rotations
    for $\delta$ equal $0.5^{\circ}$, $1.0^{\circ}$ and $3.0^{\circ}$.
    All angles are in degrees. 
}
\label{tab_betas}
\end{table}

Before proceeding to the next section, it is worth making the following observation.
The above parameter $\delta$ may characterize orientations or misorientations.  
However, errors in crystallite misorientations are a consequence of errors in orientation determination.
Assuming that 'measured' orientations 
deviate from (randomly distributed) true orientations 
by random errors of fixed magnitude $\delta_{o}$ ($0 < \delta_{o} < \pi/2$), 
the deviations $\delta_{m}$ between the error-affected 
and true misorientations are 
in the range from $0$ to $2 \delta_{o}$, and their average equals
$$
\langle \delta_{m} \rangle = 
\frac{\sin \delta_{o} -\delta_{o} \cos \delta_{o} }{\sin^2 (\delta_{o}/2)} 
 \ .
$$
This expression \correction{was derived by analytical} averaging over orientations, 
axes of orientation errors and axes of misorientations. 
The angle $\langle \delta_{m} \rangle$ is well approximated by $4 \delta_{o}/3$
up to relatively large values of $\delta_{o}$. 
Thus, in short, if random orientations are affected by errors of small magnitude $\delta_{o}$,
then the average error of misorientations has the magnitude of about \correction{$1.333 \, \delta_{o}$}.


\section{Axis deviations for 'randomly' distributed rotation errors \label{sec:distributed_delta}}

\begin{figure}[h]
	\begin{picture}(300,192)(0,0)
		\put(60,0){\resizebox{10.0 cm}{!}{\includegraphics{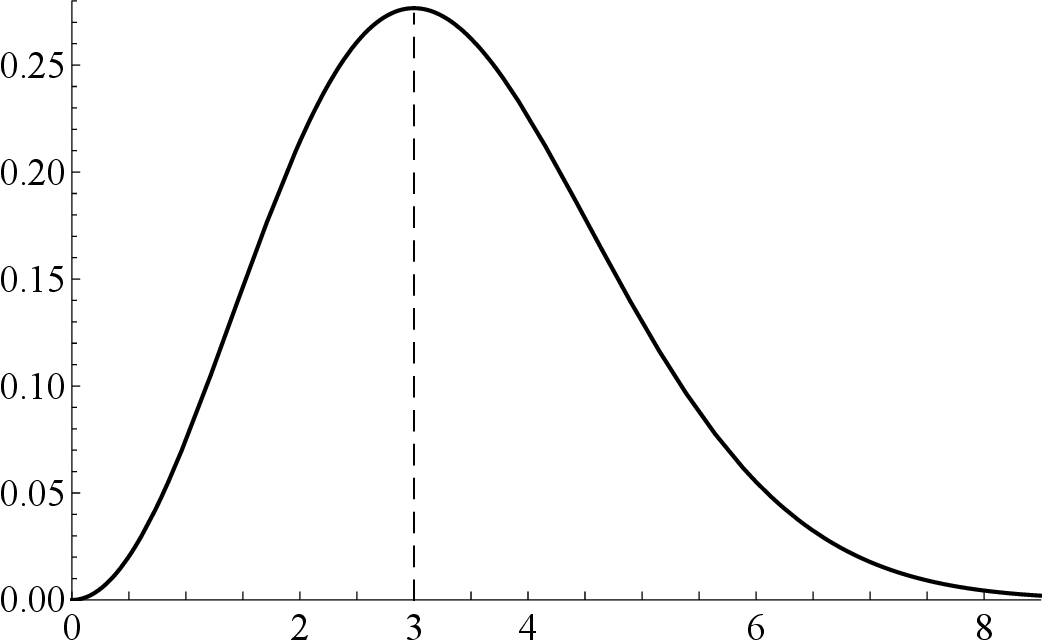}}}
		
		\put(340,0){$\delta$ [deg]}
		\put(28,173){$(\pi/180) p$}
		\put(280,153){$\widehat{\delta}=3^{\circ}$}
		\put(280,138){$\langle \delta \rangle_p = 3.39^{\circ}$}
	\end{picture}
	\vskip 0.0cm
	\caption{Example distribution $p$ of rotation angles $\delta$ 
		for rotation errors following the 'spherical' von Mises-Fisher distribution. The maximum of 
		$p$ is at $\delta=3^{\circ}$.
	}
	\label{Fig_noise_3deg}
\end{figure}
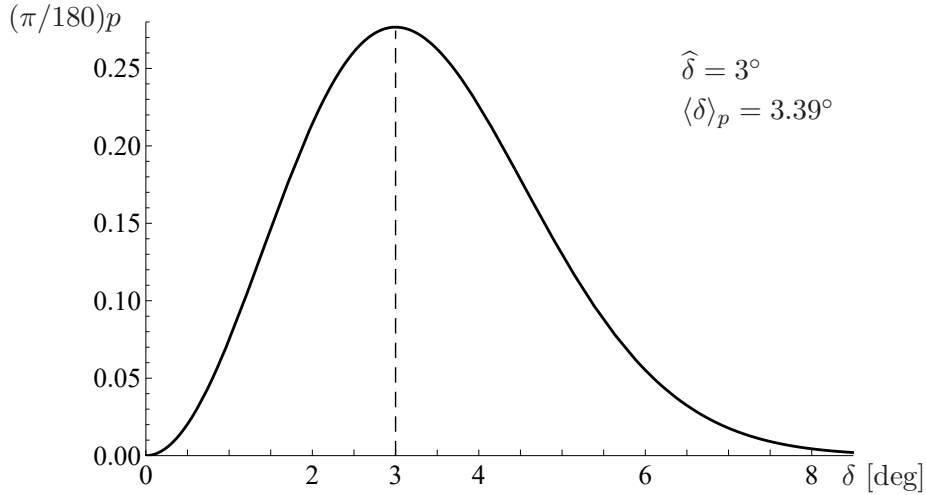

With typical experimental orientation or misorientation data, 
the magnitude $\delta$ of the error rotation
is not fixed, but it varies,
and $\langle \beta \rangle$ depends on the distribution of $\delta$.
Given little knowledge of the nature of rotation errors, their distribution must be assumed.
If the rotation errors follow the unimodal 'spherically symmetric' von Mises-Fisher distribution
 \cite{Khatri_1977,Prentice_1986}, their axes are random, and
the distribution of their angles $\delta$ is 
\bEq
p(\delta, \widehat{\delta})=N(\widehat{\delta}) \, \left(\frac{\sin(\delta/2)}{\sin(\widehat{\delta}/2)} \right)^2 \ 
\exp \left(- \left(\frac{\sin(\delta/2)}{\sin(\widehat{\delta}/2)} \right)^2 \right) \ , 
\label{eq:distrib_error_magnitudes_vMF}
\eEq
where $N(\widehat{\delta})$ is the normalization coefficient 
and $\widehat{\delta}$ ($0<\widehat{\delta}\leq \pi$) is the location of the maximum of $p$;
see Appendix B. 
The parameter $\widehat{\delta}$ uniquely determines the shape of $p$ as a function of $\delta$.
Example plot of $p$ versus $\delta$ 
is shown in Fig.~\ref{Fig_noise_3deg}.

With 
the frequency of occurrence of error magnitudes $\delta$ described by $p$, 
the  $\omega$-dependence of 
the average deviation  $\langle \beta \rangle_p$ between axes of 
the true and error-affected rotations is 
\bEq
\langle \beta \rangle_p(\omega)= \int_0^{\pi} \langle \beta \rangle(\omega, \delta) \ p(\delta, \widehat{\delta}) \ 
\mbox{d}  \delta \ . 
\label{eq:aver_beta_p}
\eEq
\correction{
To compute $\langle \beta \rangle_p(\omega)$ on needs $\langle \beta \rangle(\omega, \delta)$ 
expressed by the general formula  (\ref{eq:av_beta_1})
covering the range $0 < \delta \leq \pi$.}

\begin{figure}[h]
	\begin{picture}(300,405)(0,0)
		\put(60,200){\resizebox{10.0 cm}{!}{\includegraphics{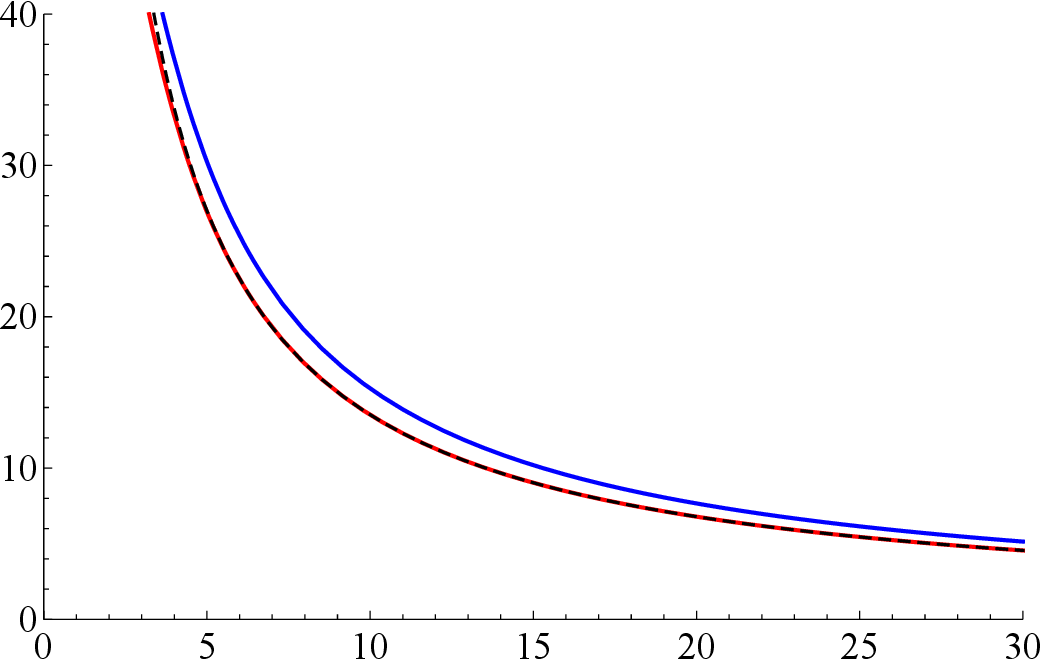}}}
		\put(60,0){\resizebox{10.0 cm}{!}{\includegraphics{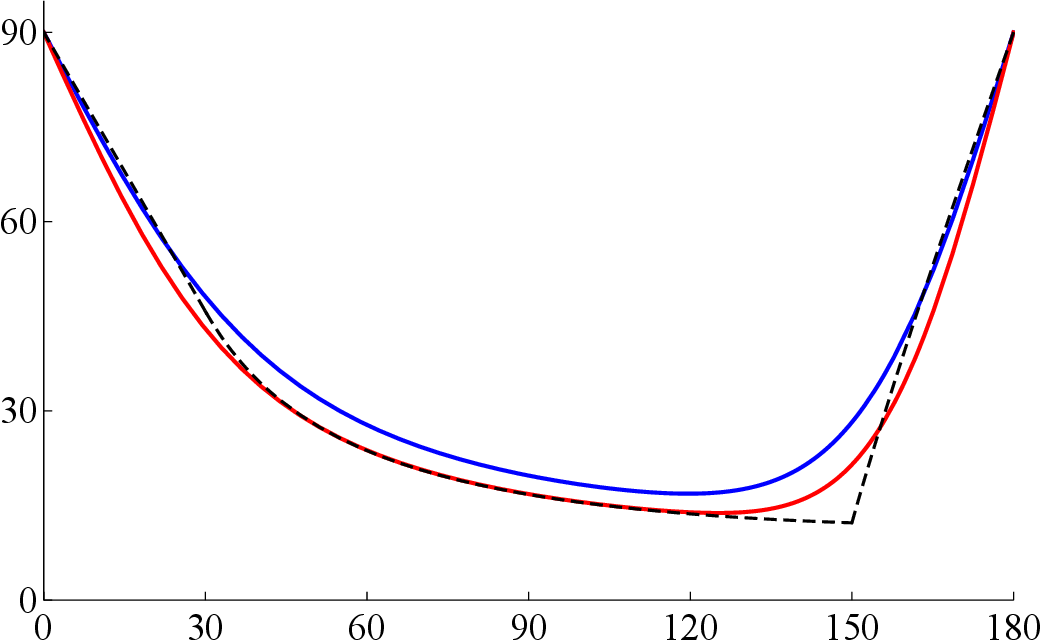}}}
		
		\put(141,296){$\widehat{\delta}=3^{\circ}$} 
		\put(153,283){$\langle \delta \rangle_p = 3.39^{\circ}$} 
		
		\put(109,262){$\langle \delta \rangle_p = 3^{\circ}$}
		\put(128,249){$\widehat{\delta}=2.66^{\circ}$} 
		\put(30,375){\textit{a}}
		\put(349,200){$\omega$ [deg]}
		\put(47,386){$\langle \beta \rangle_p$  [deg]}
		\put(30,170){\textit{b}}
		\put(349,0){$\omega$  [deg]}
        \put(47,181){$\langle \beta \rangle_p$  [deg]}
        \put(119,96){$\widehat{\delta}=30^{\circ}$} 
        \put(131,83){$\langle \delta \rangle_p = 35.06^{\circ}$} 
        \put(94,52){$\langle \delta \rangle_p =30^{\circ}$}
        \put(113,39){$\widehat{\delta}=25.94^{\circ}$} 
	\end{picture}
	\vskip 0.0cm
	\caption{The average angle  
		  between axes of the actual and error-affected rotations
		versus rotation angle $\omega$ for error magnitudes $\delta$ distributed as $p(\delta, \widehat{\delta})$
		with $\widehat{\delta}=3^{\circ}$ (blue) and 
		$\langle \delta \rangle_p=3^{\circ}$  
		(red) (\textit{a}) and 
		$\widehat{\delta}=30^{\circ}$  (blue) and 
		$\langle \delta \rangle_p=30^{\circ}$  
		(red) (\textit{b}). 
		For comparison, the graphs of $\langle \beta \rangle$ from Figs. \ref{Fig_Bate_0} and \ref{Fig_Delta_30deg} are shown in (\textit{a}) and  (\textit{b}), respectively, 
		as the dashed black curves.
	}
	\label{Fig_noise_based_3deg}
\end{figure}
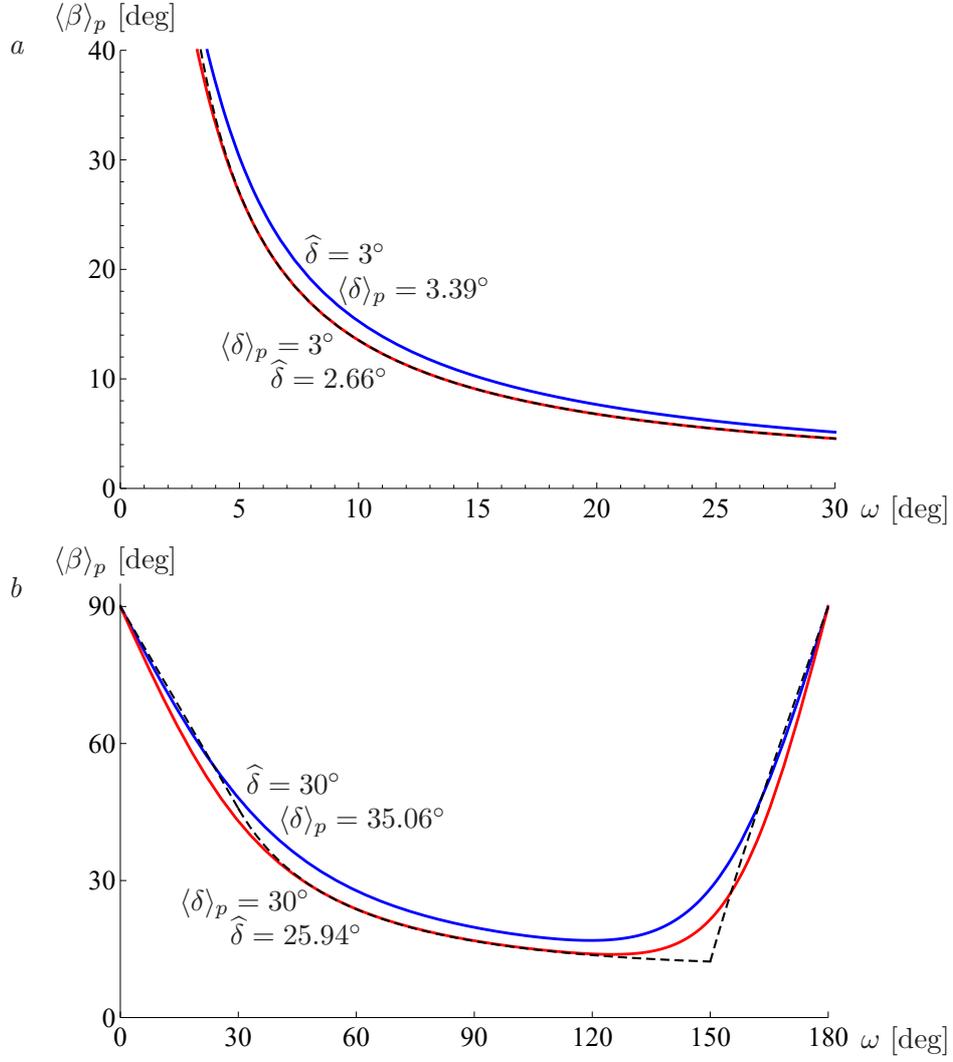

\correction{The function 
$\langle \beta \rangle_p(\omega)$}
is uniquely determined by $\widehat{\delta}$. 
Alternatively, 
the functions $p$ and $\langle \beta \rangle_p$ can be specified using the average 
$\langle \delta \rangle_p = \int_0^{\pi} \delta \, p(\delta, \widehat{\delta}) \, \mbox{d} \delta$ 
instead of  $\widehat{\delta}$.
For small $\widehat{\delta}$, the angle $\langle \delta \rangle_p$ 
is approximately given by $\langle \delta \rangle_p \approx 1.13\, \widehat{\delta}$.
The reason for introducing $\langle \delta \rangle_p$ is as follows:
In cases of interest to orientation-data analyses 
($\langle \delta \rangle_p$ of the order of $1^{\circ}$), 
with $\omega$ sufficiently distant from the ends of its range (say 
$2 \langle \delta \rangle_p < \omega < \pi - 2 \langle \delta \rangle_p$),
the function $\langle \beta \rangle_p(\omega)$
is well approximated by $\langle \beta \rangle(\omega,  \langle \delta \rangle_p  )$,  i.e., 
by $\langle \beta \rangle(\omega,  \delta  )$ given by (\ref{eq_beta_average})
with $\delta$ set at $\langle \delta \rangle_p$. 
Thus, in many practical situations, $\langle \beta \rangle_p$ can be estimated using
the analytical expression~(\ref{eq_beta_average}).
Example plots of $\langle \beta \rangle_p$  versus  $\omega$	
are shown in Fig.~\ref{Fig_noise_based_3deg}.

Clearly, with rotation errors scattered according to the von Mises-Fisher distribution
(i.e., the function that is nowhere equal to zero),
the maximal deviation $\beta_{\max}$ 
equals $\pi$.

\clearpage

\section{Deviations of axes in the case of symmetric objects \label{sec:symmetric_case}}

The analysis of Bate et al. concerned misorientations
between symmetric crystals.
However, to account for symmetry, one  needs an approach more subtle 
than that proposed in \cite{Bate_2005} and followed in section \ref{sect_no_symm} above.

Due to crystal symmetry, distinct but equivalent reference frames can be attached to the crystal, and 
its orientation can be represented by 
several different sets of rotation parameters.
In some analyzes of symmetric crystals, symmetrically equivalent 
representations of their orientations are discriminated.
For instance, this naturally occurs in dealing with sequences of orientations differing by small-angle rotations:
one of symmetrically equivalent crystal reference frames 
is selected to describe the first orientation of the sequence, 
and then the same frame is used to describe the subsequent orientations. 
The choice of the frame in the 
subsequent orientations relies on the knowledge that the angle of its rotation from the previous one is small. 
Similarly, positions of the originally selected reference 
frame can be tracked with known sample rotations; e.g., \cite{Prior_1999}. 
However, without such prior knowledge, the choice among equivalent frames is arbitrary.
This must be taken into account in determination of rotation axes.

Therefore, 
one needs to distinguish two cases.
The first involves some explicit or tacit assumptions about the rotations 
which make the choice of crystal reference frames unique. 
In this case, the symmetry does not play any role, 
and the approach of section \ref{sect_no_symm} is applicable.

The rest of this section deals with the second case where 
orientation data are not supplemented with any 
additional information or assumptions.
Symmetric crystals can be assigned reference systems
in many equivalent ways.
In effect, there are multiple rotations relating reference frames, and 
rotation parameters are ambiguous. 
To avoid the ambiguities, rotation parameters are limited to suitably defined domains.
This applies to rotations representing both orientations and misorientations.
However, certain features of misorientation domains differ from those of orientation domains. 
Therefore, orientations and misorientations are considered separately.

\vskip 0.2cm

\noindent
\textit{Orientations}\\
To reduce the arbitrariness in the choice of orientation parameters, 
data are placed in the 
so-called fundamental region (FR) such that 
each internal point of the region represents only one orientation, 
and points at the boundary of the region 
have some equivalent points (also at the boundary).\footnote{Formally, 
FR is a closure of 
a simply connected domain containing exactly one representative
of each equivalence class.}
FRs for orientations can be obtained 
by Voronoi tessellation of the rotation space based on points representing (proper)
rotations of the crystal point group \cite{Yeates_1993,Morawiec_1997,Morawiec_2004,Krakow_2017}.
A convenient choice for FR is the canonical region which contains 
only representations closest to the null rotation $\mathit{I}$,
i.e., the region is the Voronoi cell based on $\mathit{I}$.
Clearly, this definition of FR encompasses 
the domain for rotations described in section \ref{sect_Basics}.

Let all measured orientation data be in the $\mathit{I}$-based FR, and 
let $\omega_0$ be the angular distance from the point $\mathit{I}$ to the 
boundary of the FR, i.e., to a boundary point closest to $\mathit{I}$.
For $\delta$ and $\omega$ small compared to $\omega_0$, 
the character of both 
$\langle \beta \rangle(\omega,\delta)$ and $\beta_{\max}(\omega,\delta)$
is the same as that without symmetry. 
When $\omega$ exceeds $\omega_0 - \delta$, the error may move 
a rotation 
from the $\mathit{I}$-based Voronoi cell
to another cell. 
In this case, its equivalent in FR has a highly deviating axis. 
Hence, with $\omega > \omega_0 - \delta$, the angle $\beta$ may become large.

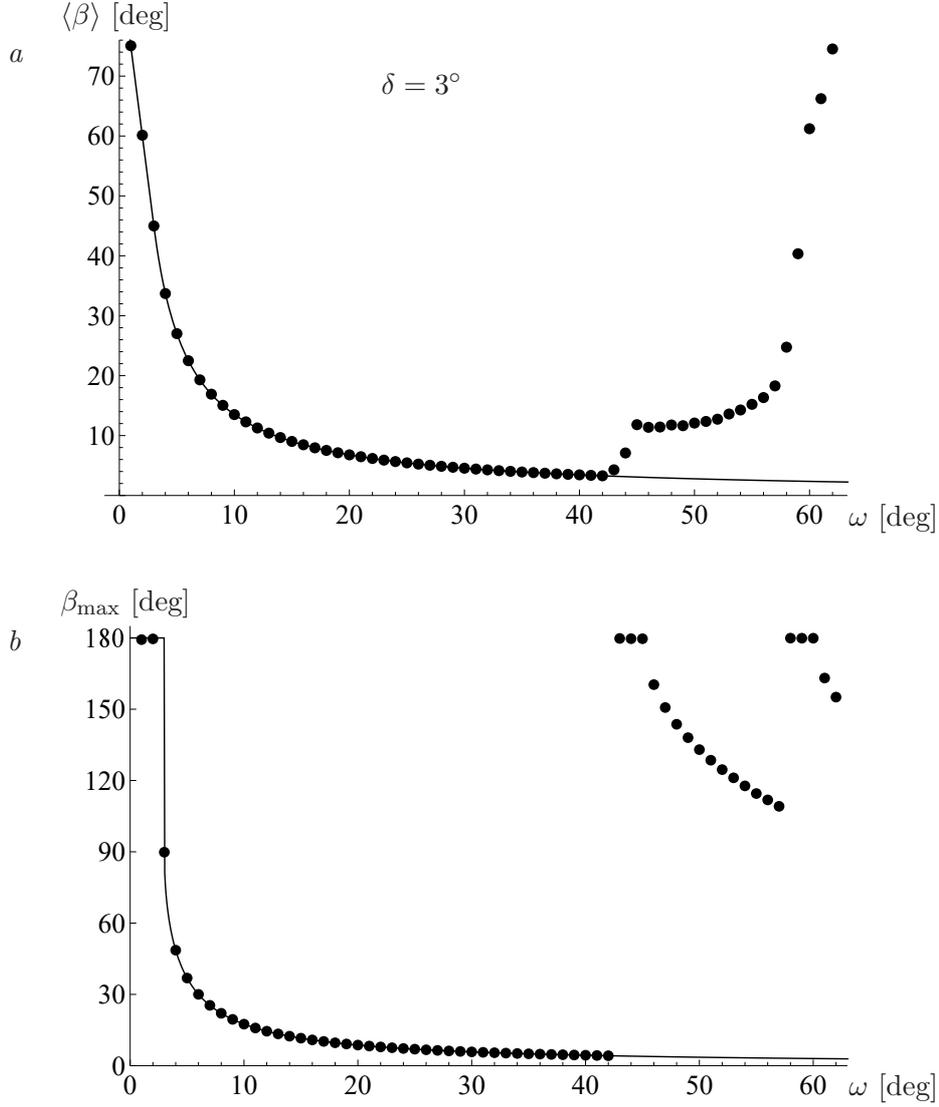
\begin{figure}
	\begin{picture}(300,410)(0,0)
		\put(60,215){\resizebox{10.0 cm}{!}{\includegraphics{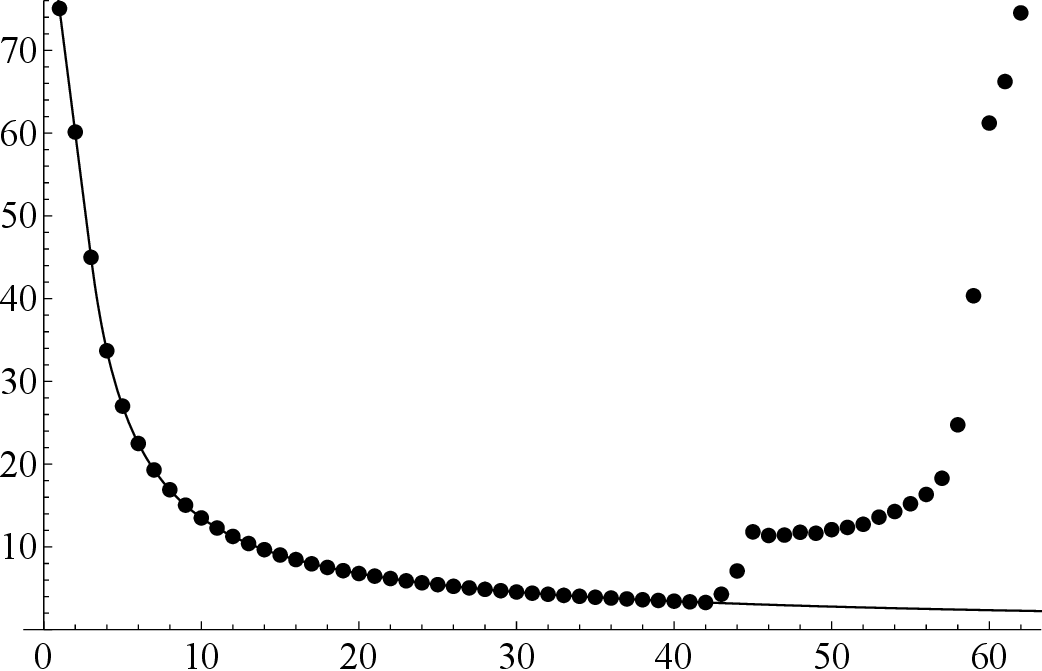}}}
		\put(60,0){\resizebox{10.0 cm}{!}{\includegraphics{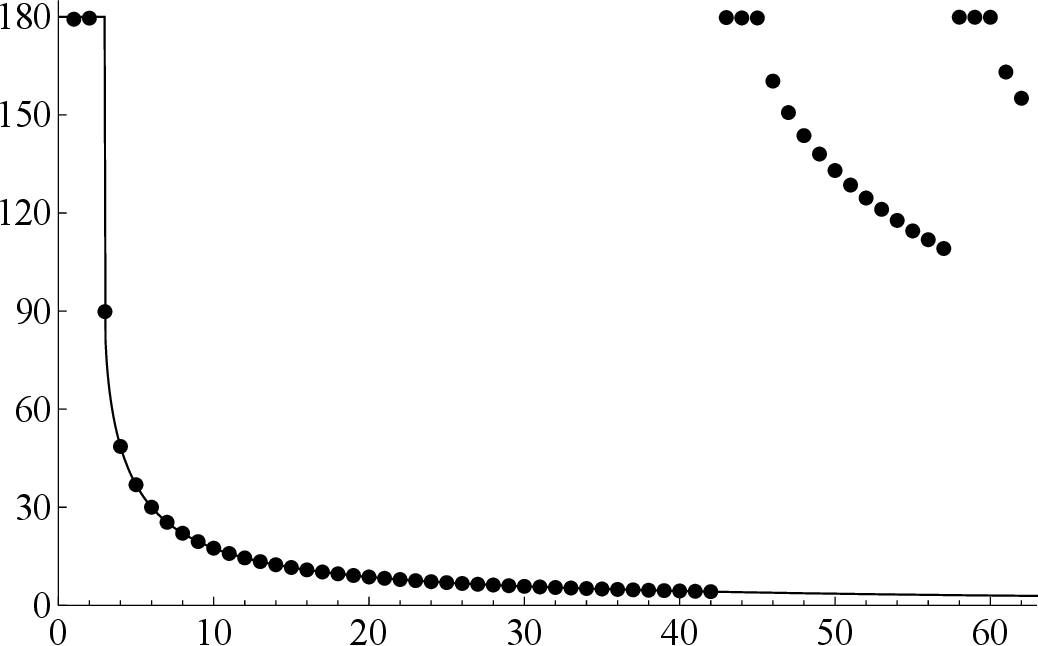}}}
        \put(30,390){\textit{a}}
		\put(345,215){$\omega$ [deg]}
		\put(50,404){$\langle \beta \rangle$   [deg]}
        \put(170,378){$\delta=3^{\circ}$}
		\put(30,168){\textit{b}}
		\put(345,0){$\omega$ [deg]}
        \put(50,183){$\beta_{\max}$  [deg]}
	\end{picture}
	\vskip 0.0cm
	\caption{
		The average  angle $\langle \beta \rangle$ 
		(\textit{a}) and the maximal  angle $\beta_{\max}$  (\textit{b})  between axes of the actual and an error-affected orientations versus rotation angle $\omega$ for the errors of magnitude $\delta=3^{\circ}$
		and cubic crystal symmetry (disks). 
		The continuous curves correspond to  $\langle \beta \rangle$ and  $\beta_{\max}$ 	
		shown in Fig.~\ref{Fig_Bate_0}. 
		Unlike the case illustrated in Fig.~\ref{Fig_Delta_30deg}, 
		where $\beta_{\max}=\pi$ for $\omega  > \pi - \delta$, here the angle 
		$\beta_{\max}$ can be smaller than $\pi$ for $\omega  > \omega_0 - \delta$. 
		The reason is that, generally, axis of  
		a rotation located just outside the boundary of FR
		is not opposite to the axis of 
		its equivalent in FR.
		This also affects the dependence of $\langle \beta \rangle$ on $\omega$ in the range
		$\omega > \omega_0 - \delta$.
	}
	\label{Fig_symm_delt_3}
\end{figure}

Unlike the case without proper symmetry 
(section \ref{sect_no_symm}), in which the average and  the maximal  values of $\beta$
are independent of \correction{the rotation axis 
$\mathbf{k}$, in the presence of symmetries, 
they depend on $\mathbf{k}$ if $\omega > \omega_0 - \delta$. 
Despite this complication $\langle \beta \rangle$ and $\beta_{\max}$ 
can be well-defined over the entire domain of $\omega$:
they represent, respectively, the average and the maximum $\beta$ over all possible directions of $\mathbf{k}$.}
The plots of these functions for $\omega > \omega_0 - \delta$
depend on the shape of the  
canonical FR which is determined by the symmetry. 
Example plots of  $\langle \beta \rangle$ and $\beta_{\max}$ versus $\omega$ 
for the cubic symmetry $m\overline{3}m$ (for which $\omega_0=\pi/4$) 
are shown in Fig.~\ref{Fig_symm_delt_3}.
Data for the plots were obtained numerically: Orientation 
having a given $\omega$ and random $\mathbf{k}$, and located in FR
was generated, 
it was perturbed by a rotation error with fixed angle $\delta$ and random axis,
the perturbed orientation was represented in FR, 
the vector $\mathbf{k}'$ along the axis was determined, 
and the angle $\beta$ between $\mathbf{k}$ and $\mathbf{k}'$ was computed. 
For a given $\omega$, these steps were repeated $10^5$ and $10^6$ times to get 
$\langle \beta \rangle$ and $\beta_{\max}$, respectively.

Summarizing, if the angle $\omega+\delta$ is smaller than $\omega_0$, 
formulas of section \ref{sect_no_symm} are applicable. 
If a rotation representing an orientation in FR 
is close to the boundary of the region,
one needs to take into consideration representations outside the region.

\vskip 0.2cm

\noindent
\textit{Misorientations}\\
As in the case of orientations, one can construct an FR for misorientations,
but unlike with orientations, there are no canonical conditions that make it unique.
The $I$-based Voronoi cell cannot be used because 
there are a number of equivalent points at exactly 
the same distance from $I$.\footnote{An additional complication is the grain exchange symmetry
which arises when crystallites are not distinguished.
With the axis/angle parameterization, the grain exchange symmetry 
comes down to the equivalence of $(\mathbf{k}, \omega)$ and $(-\mathbf{k}, \omega)$.
This symmetry is usually assumed in analyses of EBSD orientation maps.}
If the average of $\beta$ were calculated based on points in FR, 
the result would depend on the choice of the region and would vary from case to case.

Hence it follows that   
without a priori knowledge enabling the selection of a specific frame from among 
equivalent crystal reference frames,
the 
approach applicable to orientations (i.e., averaging over data in FRs) and  
figures analogous to Fig.~\ref{Fig_symm_delt_3} are of no practical 
significance
in the case of misorientations,  
and the subject of the average error in determination of misorientation axis is 
not resolved by expressions of the type (\ref{eq_beta_average}) and (\ref{eq_beta_max}). 
Therefore, when encountering the average error in determining 
the misorientation axis, 
one should be aware that the special case is considered 
in which reference systems are selected a priori 
and crystal symmetry does not play any role.

\section{Concluding remarks}

Axes of rotations together with rotation angles 
are the basic means of describing orientations and misorientations of crystals. 
Rotation axes are also a tool for studying some properties of crystalline materials.
Crystal orientations are usually determined with some errors, 
and the question is how this affects rotation axes obtained from such data.

With  $\beta$ denoting  the angle between axes of the actual rotation and its approximation,
the average $\langle \beta \rangle$ and the maximal $\beta_{\max}$  are convenient measures
of the uncertainty of the rotation axes.
The paper addresses the formal issue of determining $\langle \beta \rangle$
and  $\beta_{\max}$ 
when rotation data are affected by random errors of fixed magnitude. 
Analytical formulas for $\langle \beta \rangle$
and  $\beta_{\max}$  as functions of the rotation angle 
and the error magnitude are provided.
Additionally, a method for estimating the average of $\beta$ 
in the practical case of rotation errors following 
the von Mises-Fisher distribution is described.
These solutions are applicable in texture-related calculations 
to estimate the reliability of rotation axes.
In particular, they are crucial to assessing
the accuracy of axes of small-angle misorientations, 
which are of interest in connection with the determination of Taylor axes and slip systems.

For data obtained from crystalline materials, 
an additional factor is the crystal symmetry.
It can be ignored only 
if there are preconditions enabling 
selection of a specific reference frame from among equivalent frames.
Otherwise, 
to avoid ambiguities, data are reduced 
to fundamental regions.
In the case of orientations, there is a canonical fundamental region which consists of 
minimum-angle rotations, and  
the average and maximal $\beta$ over data in this region are well defined. 
There is no canonical choice of the fundamental region 
for misorientations, and therefore there are no simple definitions of 
the average or maximal $\beta$.

\newpage

\section*{Appendix A}
\noindent
{\bf \large  The angles $\langle \beta \rangle$ and $\beta_{\max}$ for   $0 < \omega \leq  \pi$}

\vskip 0.2cm

\noindent
To simplify comparisons with the work of Bate et al. \cite{Bate_2005}, 
most of their notation is retained. 
The expression (A4) of  \cite{Bate_2005} for the angle $\beta$ 
(between axes of the actual and error affected rotations)
can be written as
\bEq
\beta(\mathbf{h})= 
\arccos \left(\frac{\cos \left(\delta/2\right) 
	\sin \left(\omega/2\right)+ \sin \left(\delta/2\right) \cos \left(\omega/2\right) 
	(\mathbf{k}\cdot \mathbf{h}) }{\sqrt{1-\left(\cos \left(\delta/2\right) \cos \left(\omega/2\right)-\sin \left(\delta/2\right) \sin \left(\omega/2\right) 	(\mathbf{k}\cdot \mathbf{h}) \right)^2}}\right) \ , 
\label{eq_Bates_A4}
\eEq
where $\mathbf{k}$ and $\mathbf{h}$ are unit vectors along axes of the rotation $g$ and the 
error rotation $g' g^{-1}$, respectively.  
Formula (\ref{eq_Bates_A4}) 
does not account for the case 
with $\omega$ exceeding 
$\pi-\delta$, 
i.e., when the error 
affected rotation is close to a half-turn. At the half-turn, the error affected rotation
switches the axis direction.
To take this into account, one needs to include the index 
$c_s(\mathbf{h})=\mbox{sgn}\left(\cos \left(\delta/2\right) \cos \left(\omega/2\right)-
\sin \left(\delta/2\right) \sin \left(\omega/2\right) (\mathbf{k}\cdot \mathbf{h}) \right)$ 
as a factor in front of the argument of $\arccos$ in eq.(\ref{eq_Bates_A4}). 
This coefficient results from consistent application 
of formulas for composition of rotations 
and definition of the rotation axis.

Due to the homogeneity of the space of rotations,
the dependence of $\beta$ on $\mathbf{h}$ 
can be explored with any $\mathbf{k}$; this can be 
the unit vector along the $z$ axis of the Cartesian coordinate system.
With  $\mathbf{h}$ specified by spherical coordinates 
$\theta$ (polar angle) and $\phi$ (azimuthal angle), 
one has  
$\mathbf{k}\cdot \mathbf{h}=\cos \theta$.
Thus, for given $\delta$ and $\omega$, the angle $\beta$ depends only on $\theta$
\bEq
\beta(\theta)= 
\arccos \left( c_s(\theta)
\
\frac{\cos \left(\delta/2\right) 
	\sin \left(\omega/2\right)+ \sin \left(\delta/2\right) \cos \left(\omega/2\right) 
	\cos \theta }{\sqrt{1-\left(\cos \left(\delta/2\right) \cos \left(\omega/2\right)-\sin \left(\delta/2\right) \sin \left(\omega/2\right) 	\cos \theta \right)^2}}\right) \ . 
\label{eq_beta_theta}
\eEq
With random distribution of $\mathbf{h}$, the average of $\beta$ is given by the integral over the 
unit sphere $S^2$
$$
\langle \beta \rangle =
\int_{S^2} \beta(\mathbf{h}) \, \mbox{d}\mathbf{h} =
\frac{1}{4 \pi}
\int_{S^2} \beta(\theta) \sin\theta \, \mbox{d}\theta \, \mbox{d}\phi =
\frac{1}{2}
\int_{0}^{\pi} \beta(\theta) \sin\theta \, \mbox{d}\theta \ . 
$$
The integration can be carried out using software for symbolic computation.
Let 
$$
\xi_1^{\pm}(\omega,\delta) = 1 \pm \cos \left(\delta/2\right) \cos \left(\omega/2\right) \ \ \mbox{and}
$$
$$
\xi_2^{\pm}(\omega,\delta) = 
\pi -2 \arccos \left(\sin \left(\delta/2\right) \sin \left(\omega/2\right) \mp 
\xi_1^{\pm}(\omega,\delta) \ \cot \left(\delta/2\right) \cot \left(\omega/2\right) \right)
\ , 
$$
where either lower or upper sign is used concurrently on both sides. 
For 
$0 < \omega < \pi-\delta$
the dependece of the average of $\beta$ on $\omega$ and $\delta$
is 
\bEq
\langle \beta \rangle(\omega,\delta) =
\frac{1}{2} \arccos\left( \mbox{sgn}\left(\omega-\delta\right)\right) + 
\frac{\pi}{4} \ 
\frac{  \mbox{sgn}\left(\omega-\delta\right) \, 
	 \xi_1^{-}(\omega,\delta)   
	+
	 \xi_1^{+}(\omega,\delta)  
	- 2 \cos \left( \delta/2 \right)}{\sin \left(\delta/2\right) \sin \left(\omega/2\right)} \ ,
\label{eq:av_beta_1}
\eEq
and if 
$\omega \geq  \pi - \delta$, 
then 
\bEq
\langle \beta \rangle(\omega,\delta) =
\frac{1}{2} \arccos(\mbox{sgn}(\omega-\delta)) + \xi_3(\omega,\delta) + \xi_4(\omega,\delta) 
- \mbox{sgn}\left( \omega-\delta \right) \xi_5^{-}(\omega,\delta) +  \xi_5^{+}(\omega,\delta) \ ,
\label{eq:av_beta_2}
\eEq
where
$$
\xi_3(\omega,\delta) =\frac{\pi}{2} -
\frac{\cot \left(\delta/2\right)}{ \sin\left(\omega/2\right)} \ 
\arcsin\left(\cot \left(\delta/2\right) \, \cot \left(\omega/2\right)\right) \ ,
$$
$$
\xi_4(\omega,\delta) = 
\frac{1}{2} \ 
\left(\arccos\left(\frac{\cos \left(\delta/2\right)}{\sin \left(\omega/2\right)}\right)-
\arccos\left(-\frac{\cos \left(\delta/2\right)}{\sin \left(\omega/2\right)}\right)\right)
\, \cot \left(\delta/2\right) \, \cot \left(\omega/2\right) \ ,
$$
$$
\xi_5^{\pm}(\omega,\delta) =   
\frac{\xi_1^{\pm}(\omega,\delta) \  \xi_2^{\pm}(\omega,\delta)}{4 \sin(\delta/2) \sin(\omega/2)} \ .
$$
If $\omega >\delta$, then eq.(\ref{eq:av_beta_1})  reduces to simple eq.(\ref{eq_beta_average}).

It is easy to get the maximal deviation $\beta_{\max}$.
With 
$\delta < \omega < \pi - \delta$,
the condition $\mbox{d}\beta/\mbox{d}\theta=0$ for extrema of $\beta(\theta)$ 
is satisfied if 
$$
\sin \theta = 0  \ \ \mbox{or} \ \ \ 
\sin \delta \sin \left(\omega/2\right) \cos \theta + 2 \sin ^2\left(\delta/2\right) \cos \left(\omega/2\right)=0 \ . 
$$
The maximum of $\beta$ corresponds to the second case, i.e., to
$\cos \theta = -\tan \left(\delta/2\right) \cot \left(\omega/2\right)$.
Substitution of this $\cos \theta$ into (\ref{eq_beta_theta}) gives 
$$
\beta_{\max}(\omega,\delta) = 
\arccos \left(  \sqrt{\frac{\cos \delta-\cos \omega}{1 -\cos\omega}}\right) \ . 
$$
If $\omega \leq \delta$ or 
$\omega \geq  \pi - \delta$, 
the rotation affected by random error of magnitude $\delta$
may have any axis, so $\beta_{\max} = \pi$.

\section*{Appendix B}

\noindent
{\bf \large  Distribution of magnitudes of 'random' error rotations}

\vskip 0.2cm

\noindent
Random measurement errors of physical quantities represented by real-valued variables 
are often assumed to have normal 
distributions.
However, the Gausssian distribution is not suitable for orientation data.
For such data, a convenient analogue of the (trivariate) normal distribution
is the function described by 
the von Mises-Fisher distribution for special orthogonal matrices;
see, e.g., \cite{Khatri_1977,Prentice_1986}. 

\begin{figure}[t]
	\begin{picture}(300,178)(0,0)
		\put(60,0){\resizebox{10.0 cm}{!}{\includegraphics{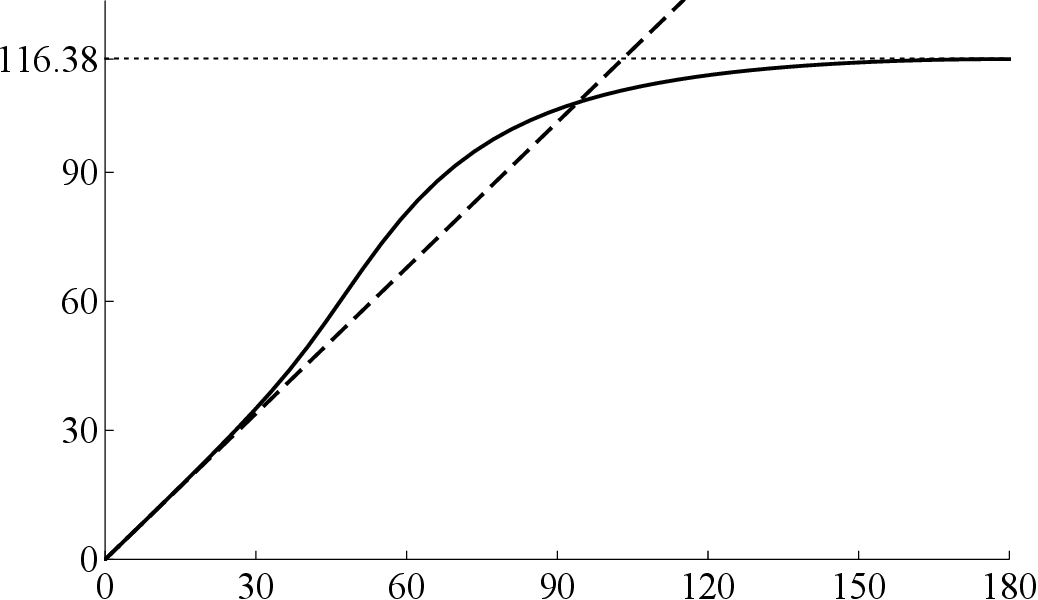}}}
		
		\put(350,0){$\widehat{\delta}$ [deg]}
		\put(70,170){$\langle \delta \rangle_p$  [deg]}
	\end{picture}
	\vskip 0.0cm
	\caption{The average $\langle \delta \rangle_p = \int_0^{\pi} \delta \, p(\delta, \widehat{\delta}) \, \mbox{d} \delta$ versus $\widehat{\delta}$ for $p$ given by (\ref{eq:distrib_error_magnitudes_vMF}). 
	The dashed line (tangent to the solid line at $0^{\circ}$) is described by $\langle \delta \rangle_p=1.13 \, \widehat{\delta}$.
	}
	\label{Delta_m_Delta_av}
\end{figure}

The simplest of the von Mises-Fisher type distributions are unimodal with 
'spherical' symmetry, i.e. they depend only on the angular distance $\delta$
between the variable 
and the mean and are independent of the axis of rotation relating these two points.
They are of the form $c(\kappa) \exp(\kappa \cos \delta)$ \cite{Khatri_1977}.
The invariant volume element on the space of rotations in axis $(\mathbf{h})$ and 
angle $(\delta)$ parameterization is 
$
(2/\pi) \sin^2(\delta/2) \, \mbox{d}\delta \, 
\mbox{d}\mathbf{h}  
$ \cite{Miles_1965}.
Thus, 
with a unimodal 'spherical' von Mises-Fisher orientation distribution, 
the distribution $p$ of the angles $\delta$ is proportional to $\sin^2(\delta/2) \exp(\kappa \cos \delta)$.
After normalization to 1, it can be expressed as 
$$
p(\delta, \widehat{\delta})=N(\widehat{\delta}) \    q(\delta, \widehat{\delta})  \,
\exp \left(-   q(\delta, \widehat{\delta})  \right) \ , 
$$
where 
$$
q(\delta, \widehat{\delta}) 
=\left( \frac{\sin(\delta/2)}{\sin(\widehat{\delta}/2)} \right)^2 \ ,
$$
the normalization coefficient 
is 
$$
N(\widehat{\delta})=\frac{1}{Q(\csc^2(\widehat{\delta}/2)/2)} \ ,
$$ 
the function $Q$ is given by
$$
Q(\kappa)=\pi \ \kappa \, e^{-\kappa} \ ( I_0(\kappa)-I_1(\kappa) ) \ , 
$$
and $I_{n}$ is the modified Bessel function of the first kind and order $n$.
The distribution $p$ can be specified using 
the location of its maximum $\widehat{\delta}$ ($0<\widehat{\delta}\leq \pi$) 
or the average 
$\langle \delta \rangle_p = \int_0^{\pi} \delta \, p(\delta, \widehat{\delta}) \, \mbox{d} \delta$.
The numerically determined one-to-one relationship of $\langle \delta \rangle_p$ to $\widehat{\delta}$ 
is shown in Fig.~\ref{Delta_m_Delta_av}.

\clearpage

\newpage

\bibliographystyle{unsrt}
\bibliography{rotationAxis.bib}

\end{document}